\documentclass[prl,twocolumn,showpacs,superscriptaddress,amsfonts,amsmath,floatfix]{revtex4}
\usepackage{graphicx}
\usepackage{psfrag}
\usepackage{epsfig}
\usepackage{pstricks}
\usepackage{epstopdf}
\newcommand{\Journal}[4]{#1 {\bf #2}, #3 (#4)}
\newcommand{\PRL}{Phys. Rev. Lett.}
\newcommand{\PRA}{Phys. Rev. A}
\newcommand{\PRB}{Phys. Rev. B}
\newcommand{\PRD}{Phys. Rev. D}

\newcommand{\lt}{\left(}
\newcommand{\rt}{\right)}
\newcommand{\lqu}{\left[}
\newcommand{\rqu}{\right]}

\newcommand{\be}{\begin{equation}}
\newcommand{\ee}{\end{equation}}
\newcommand{\ba}{\begin{eqnarray}}
\newcommand{\ea}{\end{eqnarray}}
\newcommand{\fr}{\frac}
\newcommand{\nn}{\nonumber}
\begin{document}

\title {Noise in Bose Josephson junctions: decoherence and phase relaxation}
\author{G. Ferrini}\email{giulia.ferrini@grenoble.cnrs.fr}
\affiliation{Laboratoire de Physique et Mod\'elisation des
Milieux Condens\'es, Universit\'e Joseph Fourier and CNRS, B.P. 166, 38042 Grenoble, France}
\author{D. Spehner}
\affiliation{Laboratoire de Physique et Mod\'elisation des
Milieux Condens\'es, Universit\'e Joseph Fourier and CNRS, B.P. 166, 38042 Grenoble, France}\affiliation{Institut Fourier and CNRS, Universit\'e Joseph Fourier, B.P. 74, 38402 Saint Martin d'H\`eres, France}
\author{A. Minguzzi}
\affiliation{Laboratoire de Physique et Mod\'elisation des
Milieux Condens\'es, Universit\'e Joseph Fourier and CNRS, B.P. 166, 38042 Grenoble, France}
\author{F.W.J. Hekking}
\affiliation{Laboratoire de Physique et Mod\'elisation des
Milieux Condens\'es, Universit\'e Joseph Fourier and CNRS, B.P. 166, 38042 Grenoble, France}
\begin{abstract}
Squeezed states and macroscopic superpositions of coherent states have been predicted to be 
generated dynamically in Bose Josephson junctions. We solve exactly the 
quantum dynamics of such a junction in the presence  of a classical noise coupled to the 
population-imbalance number operator (phase noise), accounting  for e.g. the experimentally relevant 
fluctuations of the magnetic field. 
We calculate the correction to the decay of the visibility induced by the noise in the non-markovian regime.
Furthermore, we predict that such a noise induces an anomalous rate of decoherence among the components of the macroscopic superpositions, which is independent on the total number of atoms,
leading to potential interferometric applications.
\end{abstract}
\maketitle

\section{I. Introduction}
It has been realized in the last decade that  an ultracold Bose gas
trapped in an optical potential offers the possibility to
manipulate coherently entangled many-body quantum states, with interesting 
applications in precision measurements and in quantum information.
For instance, spin
squeezed states and macroscopic superpositions  of coherent states are generated by the 
dynamics of a Bose Josephson Junction (BJJ)~\cite{Smerzi97,varie_gatti,noi}.
The usefulness of squeezed states
in improving phase sensitivity in interferometry has been demonstrated in a recent
experiment~\cite{Oberthaler10}. An even better sensitivity is predicted
to arise by employing macroscopic superpositions~\cite{Smerzi09}.
The presence of noise and of coupling with the environment causes decoherence and limits the experimental time for coherent manipulations.
Decoherence may  even prevent the production of certain entangled states, 
a fundamental issue in the quantum-to-classical transition~\cite{Giulini}.
It is thus important to
study the robustness of these nonclassical states in the presence of noise.
 Several sources of decoherence in BJJ like particle losses~\cite{Sinatra},
 collisions with thermal atoms~\cite{Anglin,Witthaut},
 interaction with the
electromagnetic
field~\cite{Moore06},
and random fluctuations  of the trapping potential~\cite{Khodorkovsky}  have been
identified and  analyzed theoretically.

In this work we solve the quantum dynamics of a BJJ in the presence of  a noise 
coupling linearly to the number-imbalance operator. This noise results from 
the fluctuations of the optical potential and of the magnetic field, which are, 
together with atom losses, the main sources of decoherence in the
experiments of Ref.~\cite{Oberthaler08,Oberthaler10,Treutlein10}. Our solution is exact and allows in particular to capture the decay 
of the  Ramsey fringes visibility at short 
times (non-markovian regime). 
Furthermore, it shows that the macroscopic superpositions of phase states
generated by the unitary dynamics are rather robust with respect to the noise 
considered. According to the usual scenario for decoherence~\cite{Orszag,Braun},
by increasing the intensity of the noise
these superpositions should be transformed into statistical mixtures 
of the same phase states at a noise intensity proportional to a negative power of 
the number of atoms, which characterizes here
the ``distance'' between the phase states.
For the aforementioned noise, we find that this is not the case. The typical  
noise intensity at which
the coherences between the distinct phase states 
are lost is {\it independent of the atom number} and
equal to (or even, for many-component superpositions, {\it larger than})
 the noise intensity at which phase relaxation occurs.
Phase relaxation means that  
each phase state of the superposition converges to  a mixture of 
Fock states and acquires a completely undefined phase. 
At intermediate noise, the phase has spread significantly but 
some entanglement remains due to the non vanishing coherences among the phase states of the superposition. 
We quantify this entanglement 
by computing the quantum Fisher information \cite{Smerzi09} and estimate 
the gain 
 in phase sensitivity with respect to separable states.

The paper is organized as follows. After introducing in Sec.II the two-mode approximation for the BJJ, we review in Sec.III the quenched dynamics of the junction in the absence 
of noise, leading to the formation of nonclassical states. Sec.IV presents our results on
 the effect of the noise on the  density matrix of the atoms and on the visibility of the Ramsey fringes, while Sec.V analyzes the degradation 
of the coherence of macroscopic superpositions. Finally Sec.VI offers a summary and some concluding remarks.

\section{II. Model}
We describe the BJJ by  a two-mode Hamiltonian~\cite{Milburn}
\begin{equation}
\label{eq:ham_dopo_mapping}
\hat{H}^{(0)}=\chi \hat J_z^2 -\lambda \hat J_z -2 K \hat J_x\;,
\end{equation}
where the
 angular momenta operators  $\hat J_x$,
$\hat J_y$, and
$\hat J_z$ are related to
 the annihilation operator $\hat{a}_j$ of an atom in the mode $j=1,2$ by
$\hat J_x=(\hat a^\dagger_1 \hat a_2+ \hat a^\dagger_2 \hat a_1)/2$,
$\hat J_y=-i (\hat a^\dagger_1 \hat a_2-\hat a^\dagger_2 \hat a_1)/2$, and
$\hat J_z  \equiv \hat n = (\hat a^\dagger_1 \hat a_1- \hat a^\dagger_2 \hat a_2)/2$, the latter being the number imbalance operator. 
We assume a fixed total number of atoms $N$, i.e. we do not account for atom losses. We
take $N$ to be even for simplicity, the odd case being qualitatively similar.
The Hamiltonian (\ref{eq:ham_dopo_mapping}) models both
 a single-component Bose gas trapped in a double-well potential~\cite{Oberthaler08} - external Josephson junction - 
and  a binary mixture of atoms
in distinct hyperfine states trapped in a single well~\cite{Hall98,Oberthaler10}
- internal Josephson junction. In the external BJJ the two modes $i$ 
correspond to the lowest-energy
spatial modes in each well.
For the internal BJJ, the two relevant modes are the two hyperfine states.
The first term in (\ref{eq:ham_dopo_mapping}) describes the
repulsive atom-atom interactions; for the external BJJ,
$\chi$ is the half of the sum of the interaction energies
$U_i$ in the two modes,
whereas for the internal BJJ
$\chi=(U_1+U_2)/2 - U_{12}$ also depends on the
inter-species interaction $U_{12}$. In both cases,
$\lambda=\Delta E +( N-1) (U_2-U_1)/2$ is related
to the difference $\Delta E = E_2 - E_1$
between the energies of the two modes.
The last term  in (\ref{eq:ham_dopo_mapping}) corresponds to tunelling
between the two wells or, in the internal BJJ, to a resonant laser field
coupling the two hyperfine states.
Both $\chi$ and $K$ are experimentally tunable parameters.

It is convenient to characterize a state $|\psi\rangle$ of the BJJ by its Husimi 
function
$Q (\theta,\phi) = | \langle \theta , \phi | \psi \rangle |^2$ on the 
classical phase space (the
Bloch sphere of radius $N/2$), where
\begin{equation} \label{eq-coherent_state}
|\theta, \phi \rangle=\sum_{n=-N/2}^{N/2} \left(  \begin{array} {c} N \\
 n+\frac{N}{2} \end{array}\right)^{1/2} \frac{\alpha^{n+N/2}}{(1+|\alpha|^2)^{N/2}}\, |n\rangle
\end{equation}
is a SU(2) coherent state~\cite{husimi}, $\alpha \equiv \tan(\theta/2) \exp(-i\phi)$,
and $| n \rangle$ is the Fock state satisfying $\hat J_z |n\rangle = n | n\rangle$.
For a coherent state (\ref{eq-coherent_state}) $Q$ is peaked around the vector
$N (\sin \theta \cos \phi ,\sin \theta \sin \phi , - \cos \theta)/2$,
the components of which are
the
expectation values of  $\hat{J}_x$, $\hat{J}_y$, and $\hat{J}_z$ in this state. 
In particular for a phase state, i.e. a coherent state with $\theta = \pi/2$, such 
a peak is located on the equator of the Bloch sphere.
A Fock state has a $\phi$-independent distribution, with a peak in $\theta$ at
$\theta = \arccos (- 2 n/N)$.

\section{III. Dynamics in the absence of noise}
\begin{figure}
\begin{minipage}{.96\columnwidth}
\includegraphics*[width=\columnwidth]{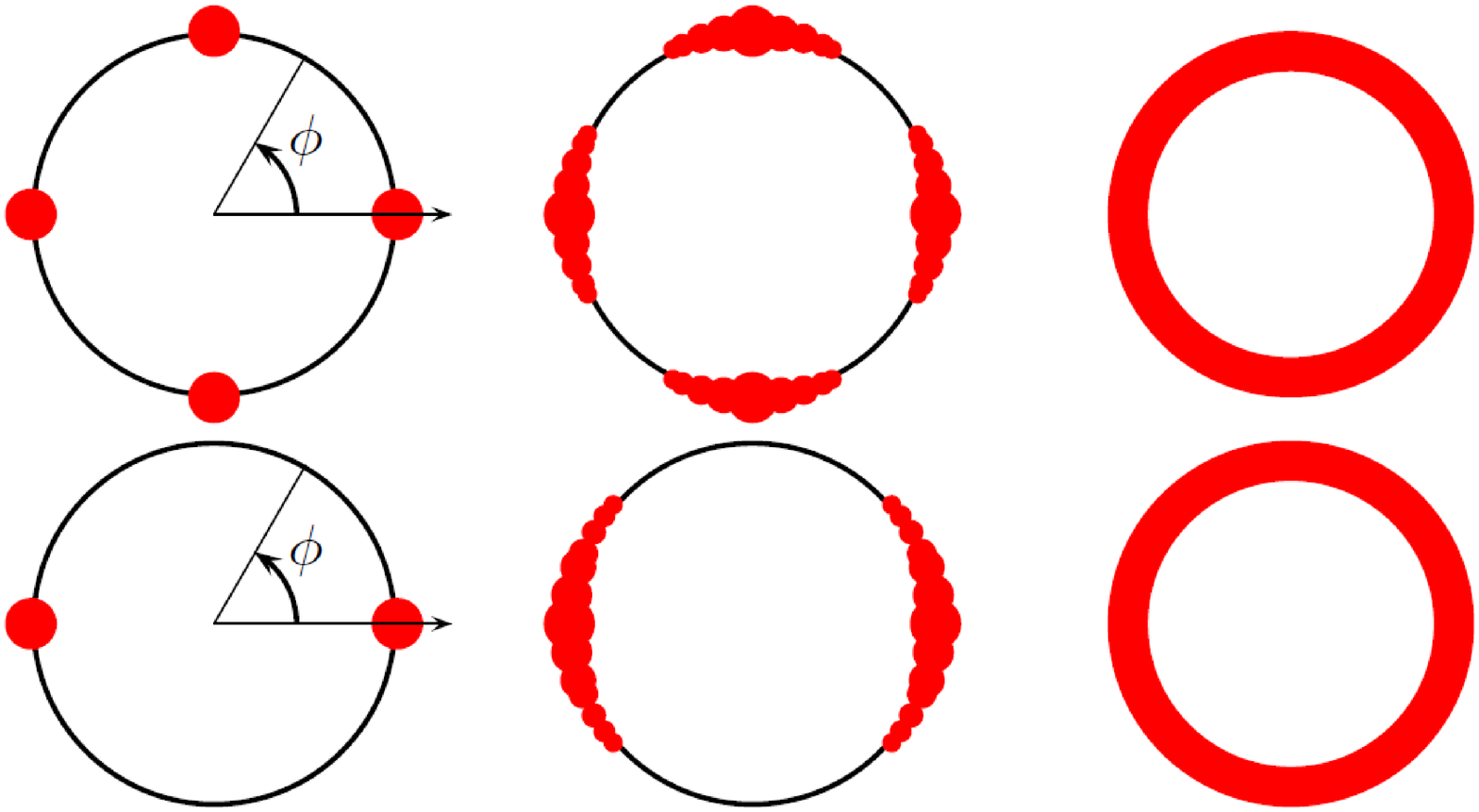}
\end{minipage}
\begin{minipage}{.32\columnwidth}
\includegraphics*[width=\columnwidth]{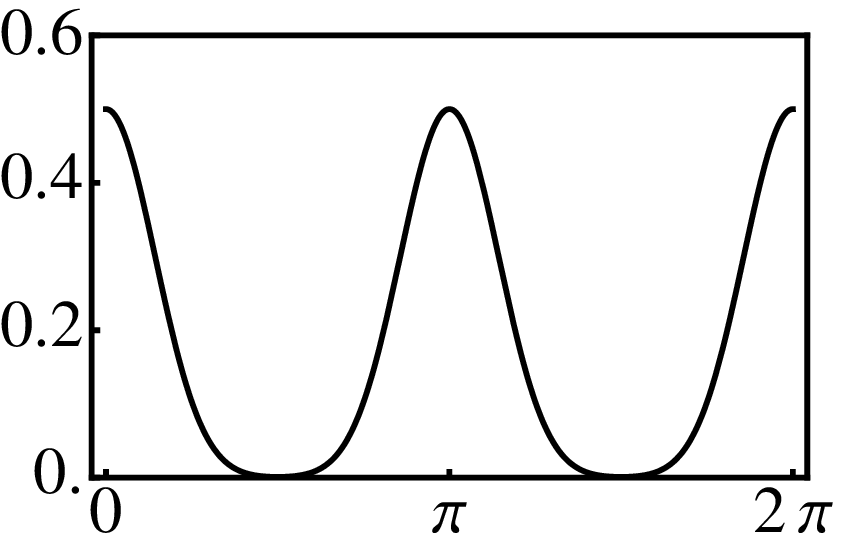}
\end{minipage}
\hfill
\begin{minipage}{.32\columnwidth}
\includegraphics*[width=\columnwidth]{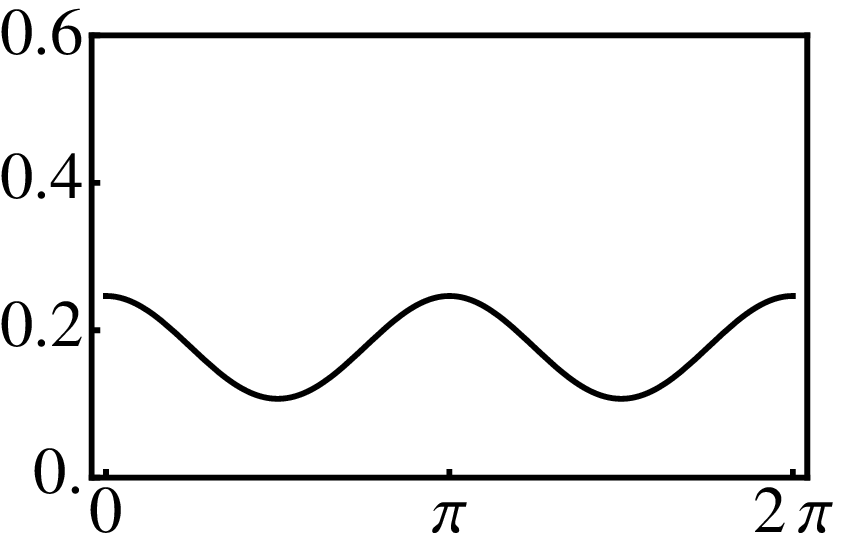}
\end{minipage}
\hfill
\begin{minipage}{.32\columnwidth}
\includegraphics*[width=\columnwidth]{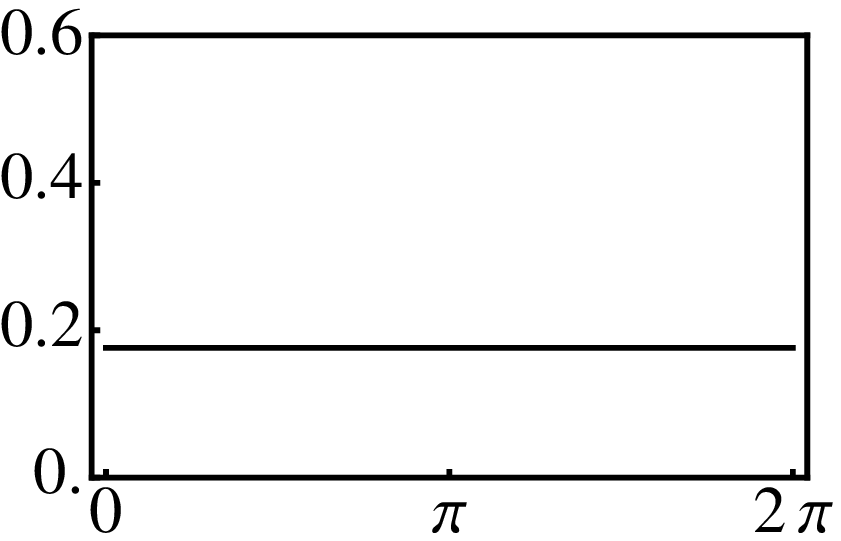}
\end{minipage}
\caption{(Color online) Phase relaxation of the $q=4$ and $q=2$ macroscopic superpositions
in the presence of noise sketched along the equator $\theta=\pi/2$ of the Bloch sphere.
Top panels: $q=4$ ($t_4=T/8$) and $a_4 = 0$, $0.64$, $2.05$ (from left to right).
Middle panels: $q=2$ ($t_2=T/4$) for the same
noise intensities $\int_0^\infty d \tau h (\tau)$
in the Markov regime ($a_2 =0$, $0.9$, $2.9$). The circle sizes illustrate qualitatively the phase distribution $f(\phi,t_{2,4})$.
For intermediate noise (middle column),
the superposition is closer to the steady state (last column)
for $q=4$ than for $q=2$.
Bottom panels: Husimi distribution $Q(\theta = \pi/2,\phi)$ for $q=2$ for the same values of $a_2$.
Here $\overline{\lambda}=0$ and $N = 10$.
}
\label{fig:sketch_rot}
\end{figure}

 In the absence of noise, let us describe the quenched dynamics of the BJJ induced by 
a sudden switch off of the Josephson coupling $K$  in (\ref{eq:ham_dopo_mapping}) at 
time $t=0$. We take $\lambda=0$ for simplicity. Initially, the BJJ is in the
 phase state with $\phi=0$
(i.e., $\alpha=1$). This state is
 the ground state of the Hamiltonian (\ref{eq:ham_dopo_mapping})
in the regime  $K N\gg \chi$ where tunelling dominates interactions.
In the internal BJJ, it can be produced by applying
a short $\pi/2$-pulse to the atoms initially in the lower level.
Under the effect of the quench the phase
starts diffusing along the equator of the Bloch sphere.
The visibility  of Ramsey fringes~\cite{Wineland}   
at time $t>0$ reads~\cite{Sinatra}
\be
\label{eq:visibility}
\nu^{(0)}(t) \equiv \frac{2}{N} \langle \hat{J}_x^{(0)}(t) \rangle 
 =   \cos^{N - 1} \lt \chi t   \rt.
\ee
At small times, the BJJ is in 
a squeezed state~\cite{Kitagawa93}.
Later on,
it returns to the initial state $|\alpha=1\rangle$ at  the revival time $T \equiv 2 \pi/\chi$;
at intermediate times $t_q \equiv T/(2q)$ it
is in a superposition
$|\psi^{(0)} (t_q)\rangle = u_0 \sum_{k = 0}^{q -1} c_k |e^{-i2 \pi k/q} \rangle $
of $q$ phase states \cite{varie_gatti,noi} with
$|u_0|^2 = 1/q$ and $c_k = e^{i \pi k (k + N)/q}$ (we have taken $q$ even).
By~(\ref{eq-coherent_state}), the matrix elements of the density matrix
$\hat{\rho}^{(0)}(t_q) = |\psi^{(0)} (t_q)\rangle \langle \psi^{(0)} (t_q) |$ 
in the Fock basis are the sum  over all $k,k'=0,\cdots,q-1$ of
\ba
\label{eq:dm_general_Fock_0}
&&
\langle n  | \hat{\rho}^{(0)}_{kk'} (t_q) |n' \rangle
 =
  \fr{1}{q} \fr{1}{2^N} {N \choose n+ \frac{N}{2}}^{\fr{1}{2}}
   {N \choose n'+\frac{N}{2}}^{\fr{1}{2}} \nn  \\
 && \qquad \qquad \qquad \times  e^{- 2 i \pi ( k n - k' n' )/q}
e^{ i \pi (k^2 - k'^2)/q}
\ea
with $
\hat{\rho}^{(0)}_{kk'} (t_q) =
q^{-1} c_k c_{k'}^*|e^{-i 2 \pi k/q} \rangle \langle e^{-i 2 \pi k'/q}  |$.
Since the dynamics does not couple the two
modes, $\hat J_z$ is a constant of motion. 
Thus $\langle n | \hat{\rho}^{(0)} (t) | n \rangle$  is constant in time and
equal to $P_{\alpha = 1}(n) = 2^{-N} {N \choose n+ N/2}$.
In order to address later on 
the decoherence and phase relaxation of the superpositions of phase states,  
we decompose 
$\hat{\rho}^{(0)}(t_q)$ as
\be
\label{eq:dm_0}
\hat{\rho}^{(0)} (t_q)
 =  \displaystyle \sum_{k = 0}^{q -1} \hat{\rho}^{(0)}_{kk} (t_q)
  +  \sum_{k \neq k' = 0}^{q -1} \hat{\rho}^{(0)}_{kk'} (t_q)\,.
\ee
The first sum in (\ref{eq:dm_0}), which we will refer to 
as the ``diagonal part'' $\hat{\rho}^{(0)}_{\text{d}} (t_q)$, is a statistical mixture of phase states.
It is mainly responsible for the structure of the phase profile given by the Husimi
distribution~\cite{Ferrini_2}.
{ The second sum in (\ref{eq:dm_0}), to be referred below as the ``off-diagonal part'' $\hat{\rho}^{(0)}_{\text{od}} (t_q)$,
accounts for quantum correlations   and interference effects, such as, for example, fringes in the eigenvalue probability distributions
of $\hat{J}_x$ and $\hat{J}_y$~\cite{Ferrini_2}.

\section{IV. Dynamics in the presence of phase noise}
We now account for the effect of noise by considering the Hamiltonian
\begin{equation}
\hat{H}(t) = \chi \hat{J_z}^2- \lambda(t) \hat J_z
\end{equation}
 where $ \lambda(t)$ is a classical stochastic process.
Since $ [\hat{H}(t),\hat{J}_z]=0$ at all times, $\hat{J}_z$ is conserved
as in the noiseless case.
Neglecting the fluctuations of
$U_i$ (which seems justified in the experiments), the fluctuations of
$\lambda$ are  equal to those of $\Delta E$ and are independent of $N$.
For a given realization of the process $\lambda$, the Schr\"odinger-evolved state
is obtained from the state
$|\psi^{(0)} (t) \rangle = e^{-i \chi \hat{J}_z^2 t} |\alpha=1 \rangle$
in the absence of noise
through a  rigid rotation around the $z$-axis
by a random angle $\phi (t) \equiv - \int_0^{t}{  d\tau \lambda(\tau)}$, i.e.,
$
|\psi (t) \rangle = e^{-i \phi(t)  \hat{J}_z }|\psi^{(0)} (t) \rangle
$.
The phase $\phi$ has a distribution
$f(\phi,t)  =
 \int d P \lqu \lambda \rqu  \delta (\phi(t) - \phi )$
where $P \lqu \lambda \rqu$ is the probability distribution of the process $\lambda$.
Averaging over all realizations of $\lambda$
leads to the density matrix
$\hat{\rho}(t) = \int d P \lqu \lambda \rqu |\psi (t) \rangle \langle \psi (t)| $.
This is the analog of tracing out the bath degrees of freedom in models of systems
coupled to quantum baths.
We obtain
\be
\label{eq:density_matrix_8}
\hat{\rho}(t)  =
\int_{- \infty}^{\infty}
d\phi \, f(\phi, t) \, e^{-i \phi  \hat{J}_z }
\hat{\rho}^{(0)}(t) e^{i \phi  \hat{J}_z }
\ee
where $\hat{\rho}^{(0)}(t) = |\psi^{(0)} (t) \rangle \langle \psi^{(0)} (t) |$ is 
the density matrix in the absence of noise.
By projecting Eq.(\ref{eq:density_matrix_8}}) over the Fock basis we get 
\ba
\langle n | \hat{\rho}(t) | n' \rangle &=& \int_{-\infty}^\infty 
d \phi f(\phi,t) e^{- i \phi (n - n')} \langle n| \hat{\rho}^{(0)}(t) | n'\rangle \nn \\
&=& \tilde{f}(n' - n,t) \langle n| \hat{\rho}^{(0)}(t) | n' \rangle
\ea
where $\tilde{f}(m,t) = \int_{-\infty}^\infty d \phi f(\phi,t) e^{ i m \phi} = \overline{e^{ i m \phi(t)} }$ 
is the Fourier transform of $f(\phi,t)$ with respect to $\phi$ and the overline denotes the average over the realizations of the noise $\lambda$ according to the 
probability distribution $P \lqu \lambda \rqu$.
To be specific, let us consider a gaussian noise. Then $\tilde{f}(m,t) = e^{-a^2(t) m^2/2} e^{-i \overline{\lambda} t m}$,
where the variance $a^2(t)$ is given in terms of the  noise  correlation function
$h(\tau - \tau') = \overline{\lambda(\tau) \lambda(\tau') }-\overline{\lambda}^2=
\overline{\Delta E (\tau)\Delta E (\tau')} - \overline{\Delta E}^2$ 
by $a^2(t) = \int_0^{t} d\tau \int_0^{t}
  d\tau' h(\tau - \tau') = 2 \int_0^{t} d\tau \int_0^{\tau}
  d u h(u)$. (Note that $h$ depends on the time difference $\tau - \tau'$  by the 
stationarity of the process, which also implies 
$\overline{\lambda(t)}= \overline{\lambda(0)}\equiv \overline{\lambda}$.)
This yields
\be
\label{eq:density_matrix_9}
\hspace*{-0.2cm}
\langle n | \hat{\rho}(t) | n ' \rangle
=
e^{-\fr{a^2(t) (n - n')^2}{2}} e^{i \overline{\lambda} t (n-n' )}
\langle n | \hat{\rho}^{(0)}(t) | n' \rangle
.
\ee
The effect of the noise
is to spread 
the noiseless evolution $\hat{\rho}^{(0)}(t) $ along the equator of
the Bloch sphere by the amount $a(t)$; in this sense, it is a pure-dephasing noise.
Since our result (\ref{eq:density_matrix_9}) does neither rely on a perturbative approach nor on a Markov approximation,
it is valid also for strong noise and at short times. The variance $a^2(t)$ 
does not depend on $N$ and completely characterizes the effect of the noise on the BJJ.
It is given by
\be
\label{eq-a^2(t)}
a^2(t) \simeq
\begin{cases}
h (0)\, t^2
& \text{if $t \leq t_c\,$ (small time)}
\\
 2 \int_{0}^{\infty} d \tau  h ( \tau)  \, t
&   \text{if $t \gg T_c$ (Markov)}
\end{cases}
\ee
where we have introduced the noise  time scales $t_c$ and $T_c$, 
$t_c$ being the largest time such that
 $h (\tau) \simeq h (0) = \delta \lambda(0)^2$ for
$|\tau | \leq t_c$ and $T_c$ the smallest time such that
$h (\tau) \simeq 0$ for $\tau \geq T_c$.
Eq.~(\ref{eq:density_matrix_9}) shows how the  noise suppresses the off-diagonal elements 
of the density matrix in the Fock basis. At long times
$t \gg (\int_0^\infty d \tau h(\tau) )^{-1}$
the state of the BJJ converges to a statistical mixture of Fock states
with the same probabilities as  the initial state,
\be
\label{eq:result_3}
\hat{\rho} (\infty)
 =
  \hspace{-0.30cm}   \sum_{n=-N/2}^{N/2}  \hspace{-0.30cm}
 P_{\alpha = 1}(n) \bigl| n \rangle \langle n \bigr|
\hspace{-0.1mm} = \hspace{-0.1mm}
\int_0^{2 \pi} \hspace{-0.1mm}  \frac{d \phi }{2 \pi}   \bigl| \frac{\pi}{2}, \phi \bigr\rangle
\bigl\langle \frac{\pi}{2}, \phi \bigr| .
\ee
The last equality is
obtained from  (\ref{eq-coherent_state}). It means that
at large times
the phase $\phi$ is uniformly spread on $[0,2\pi]$, as is the case for Fock states
(Fig.\ref{fig:sketch_rot}, right panels).

For a non-gaussian noise,  the two exponentials in the
right-hand side of (\ref{eq:density_matrix_9}) coincide with 
the cumulant expansion 
of $\tilde{f} (m,t) = \overline{e^{i m \phi (t)}}$ up to the second 
cumulant~\cite{vanKampen} and
higher cumulants yields extra factors $e^{b_p (t) (n-n')^p}$,
$p=3,4,\ldots$. For  small times $t \leq t_c$, these factors are close to unity and
the
right-hand side of (\ref{eq:density_matrix_9}) still gives 
a good approximation of the matrix elements of $\hat{\rho}(t)$.

\begin{figure}
\begin{minipage}{\columnwidth}
\includegraphics[width=0.8\columnwidth]{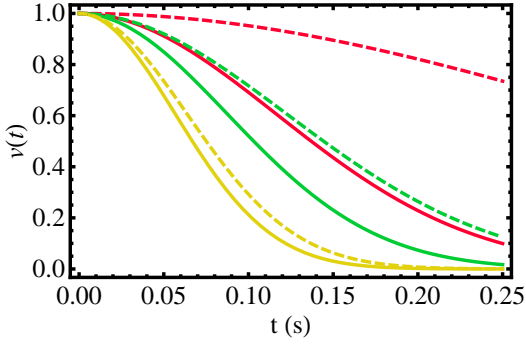}
\end{minipage}
\caption{(Color online) { Visibility $\nu(t)$ as a function of time (in units of seconds) for $\chi=  \pi \cdot 0.05$Hz, 
$\pi \cdot 0.13$Hz, $ \pi \cdot 0.25$Hz (from top to bottom), $N = 400$. 
Solid lines: decay of $\nu(t)$ in Eq.(\ref{eq:decoherence_coherence_factor}) in the 
limit $\chi t \ll 1$ and $\overline{\lambda} t\ll 1$ with $a^2(t) = h(0) t^2$ and 
$h(0)^{1/2} = 8$Hz. Dashed lines: decay of $\nu^{(0)}(t)$ under the unitary evolution only. 
For small values of the interactions the decay is mainly due to the noise.}
}
\label{fig:visibility}
\end{figure}

Under the effect of the noise, the visibility (\ref{eq:visibility})
acquires an additional decaying factor due to the above-mentioned phase spreading.
Indeed, one easily obtains from (\ref{eq:density_matrix_9})
\be
\label{eq:decoherence_coherence_factor}
\hspace*{-0.2cm}
\nu(t) =  \frac{2}{N} {\rm tr} [ \hat{\rho}(t) \hat{J}_x ] 
= e^{-\frac{a^2(t)}{2}} \cos{\lt \overline{\lambda} t \rt} \nu^{(0)}(t).
\ee
As  found e.g.
in superconducting circuits~\cite{Ithier} and in quantum dots~\cite{Vagov},
the dephasing factor $e^{-a^2(t)/2}$ displays a Gaussian decay
at short times $t \leq t_c$, corresponding to the universal regime of Ref.~\cite{Braun}, 
and an  exponential decay at long times $t \gg T_c$
(Markov regime).
A Gaussian 
decay of the visibility (\ref{eq:decoherence_coherence_factor}) has been observed 
experimentally in the internal BJJ 
at small values of the interactions $\chi$~\cite{Gross_thesis}. 
This indicates that the experiment was performed in the small-time, non-markovian regime. 
The effect of the noise on the visibility decay in this regime is shown in 
Fig.\ref{fig:visibility}, for experimentally relevant parameters~\cite{Gross_thesis}.
 A direct comparison with the experiment requires to include  in the model atom losses.

\section{V. Decoherence and relaxation of macroscopic superpositions}

Let us then study the impact of decoherence and relaxation 
on a macroscopic superposition 
at the fixed time $t_q $ as a function of 
the noise intensity. We assume that $q$ is even. 
Using (\ref{eq:density_matrix_9}) and the decomposition (\ref{eq:dm_0}) of the density matrix into diagonal and off-diagonal parts, 
and setting $a_q \equiv a(t_q)$, one gets
\be \label{eq:density_matrix_10}
 \langle n |  \hat{\rho}_{\text{d,od}}(t_q) | n' \rangle
 =
e^{-\fr{a^2_q (n - n')^2}{2}}  \langle n|  \hat{\rho}_{\text{d,od}}^{(0)}(t_q) | n'\rangle
\ee
up to a phase factor irrelevant for decoherence.
 Since the effect of the noise in Eq.(\ref{eq:density_matrix_10}) factorizes out let 
us concentrate on the structure of the density matrix in the absence of noise. 
By~(\ref{eq:dm_general_Fock_0}) and the definition following Eq.(\ref{eq:dm_0}) we have  $\langle n | \hat{\rho}_{\rm d}^{(0)}(t_q )| n' \rangle \propto
\sum_{k=0}^{q-1} e^{-\frac{2 i \pi}{q} k (n-n')}=0 $ if $n' \not= n$ modulo $q$
and 
$$\langle n | \hat{\rho}_{\rm d}^{(0)}(t_q )| n' \rangle= \frac{1}{2^{N}} 
 \left(  \begin{array} {c} N \\ n+\frac{N}{2} \end{array}\right)^{1/2}  
\left(  \begin{array} {c} N \\
 n'+\frac{N}{2} \end{array}\right)^{1/2}
$$
 otherwise.
Therefore
\be
\label{casi}
\langle n | \hat{\rho}_{\rm d}^{(0)}(t_q )| n' \rangle = 
\begin{cases}
\langle n | \hat{\rho} (0 )| n' \rangle
& \text{if $n' = n$ modulo $q$}
\\
0
&   \text{if $n' \not= n$ modulo $q$}
\end{cases}
\ee
and 
\be
\label{casi2}
\hspace{-0.25cm} \langle n | \hat{\rho}_{\rm od}^{(0)}(t_q )| n' \rangle = 
\begin{cases}
0
& \hspace{-0.25cm} \text{if $n' = n$ modulo $q$}
\\
e^{i \frac{\pi}{q} ( {n'}^2 -  n^2)} \langle n | \hat{\rho} (0 )| n' \rangle
& \hspace{-0.25cm}  \text{if $n' \not= n$ modulo $q$}
\end{cases}
\ee
The last equality follows from (\ref{casi}) and from the identity
$ \hat{\rho}^{(0)} (t_q )  
= e^{-i t_q \chi \hat{J}_z^2} \hat{\rho} (0 ) e^{i t_q \chi \hat{J}_z^2}$.
In particular, we get
$\langle n | \hat{\rho}_{\text{od}}^{(0)}(t_q) | n \pm 1 \rangle \not= 0$ 
from  (\ref{casi2}) (since $q\geq 2$ is even).

\begin{figure}
\begin{minipage}{.32\columnwidth}
\includegraphics*[width=\columnwidth]{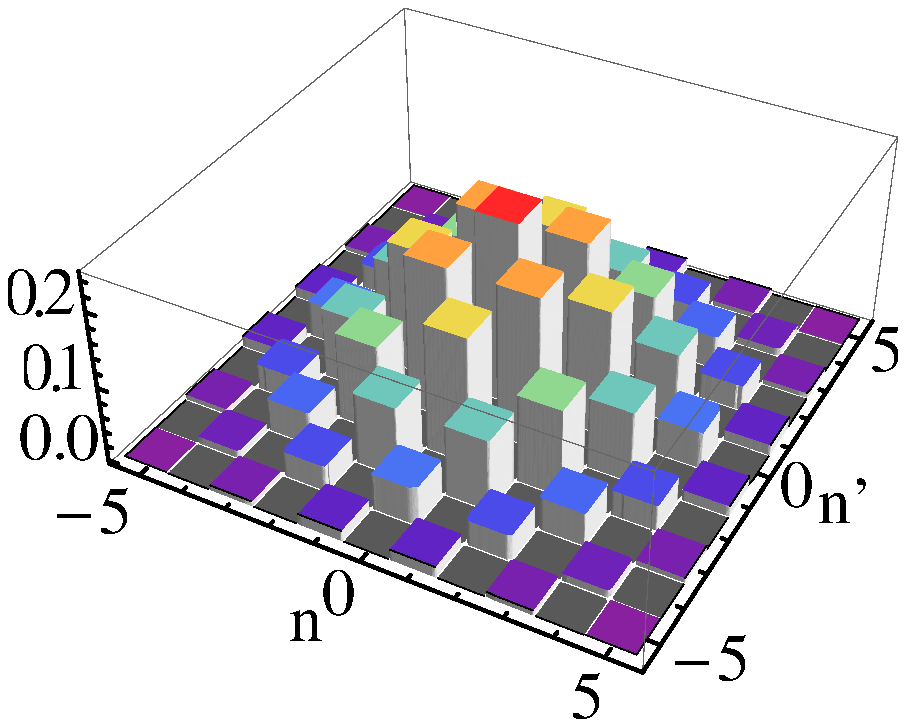}
\end{minipage}
\hfill
\begin{minipage}{.32\columnwidth}
\includegraphics*[width=\columnwidth]{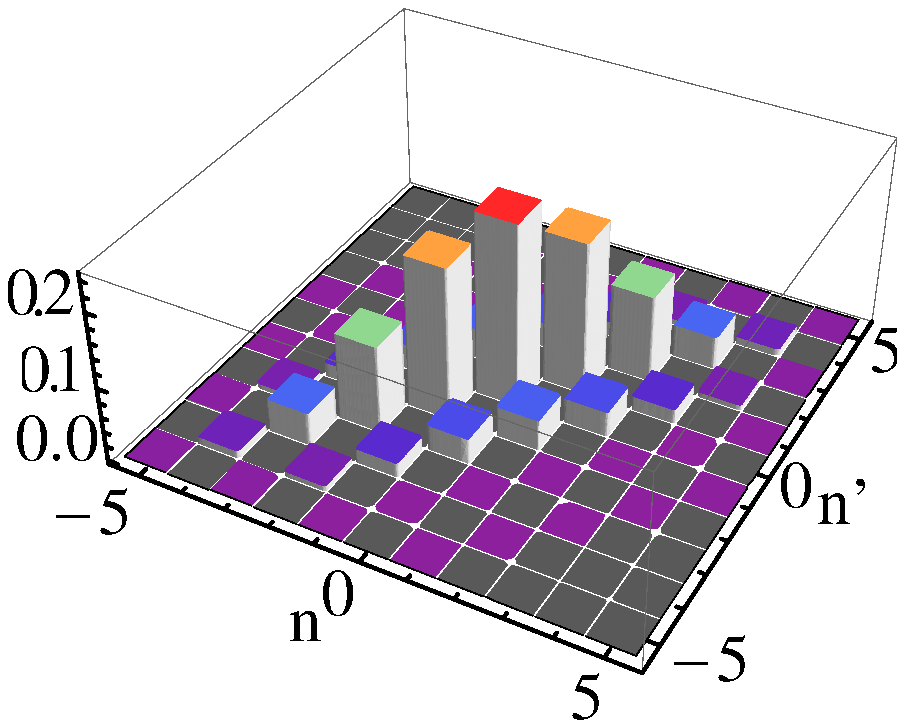}
\end{minipage}
\hfill
\begin{minipage}{.32\columnwidth}
\includegraphics*[width=\columnwidth]{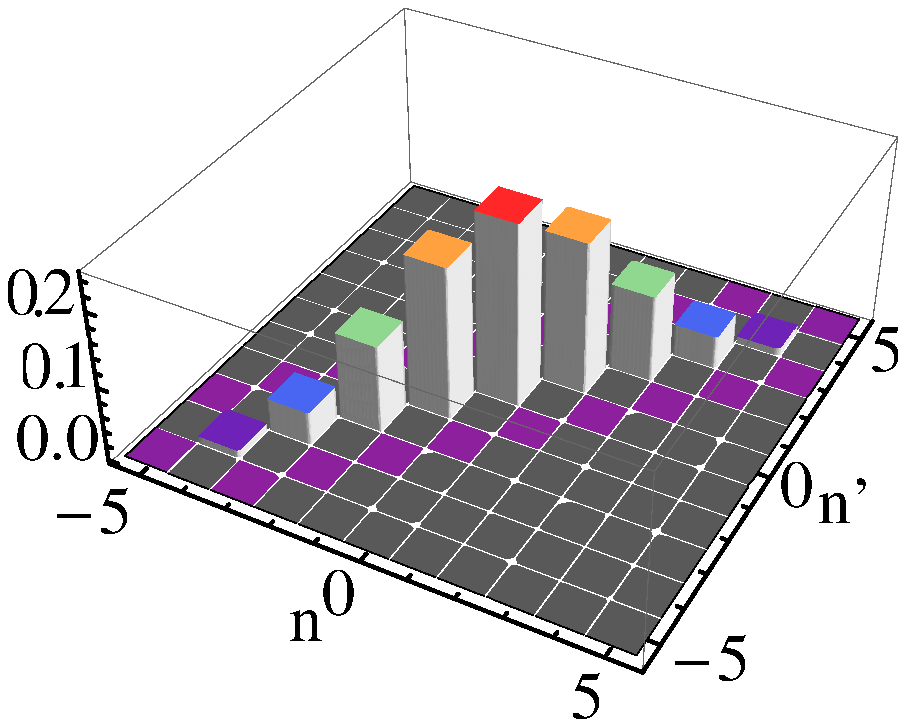}
\end{minipage}
\begin{minipage}{.32\columnwidth}
\includegraphics*[width=\columnwidth]{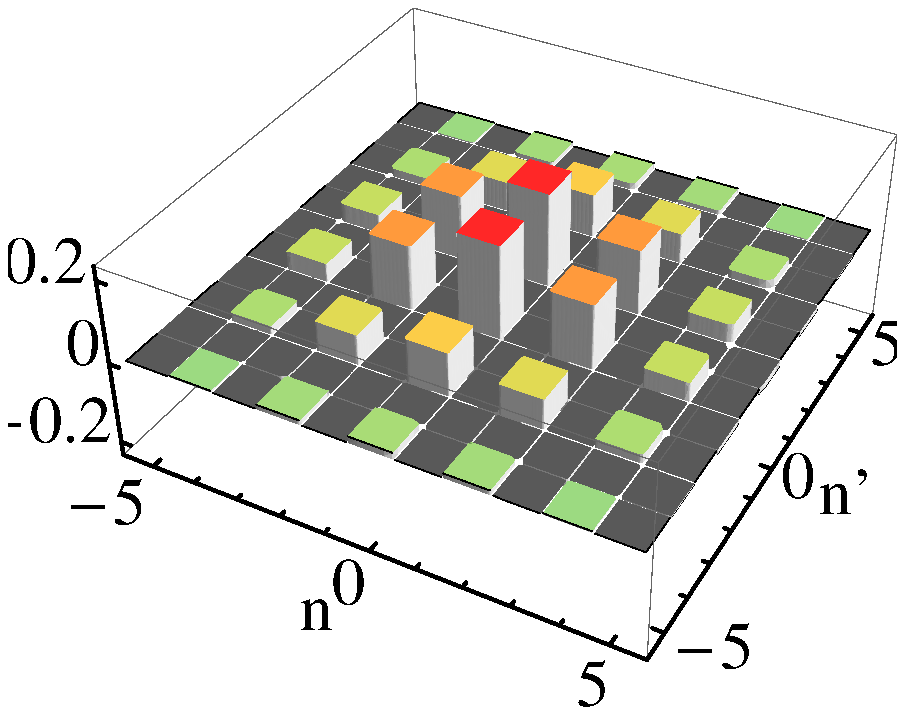}
\end{minipage}
\hfill
\begin{minipage}{.32\columnwidth}
\includegraphics*[width=\columnwidth]{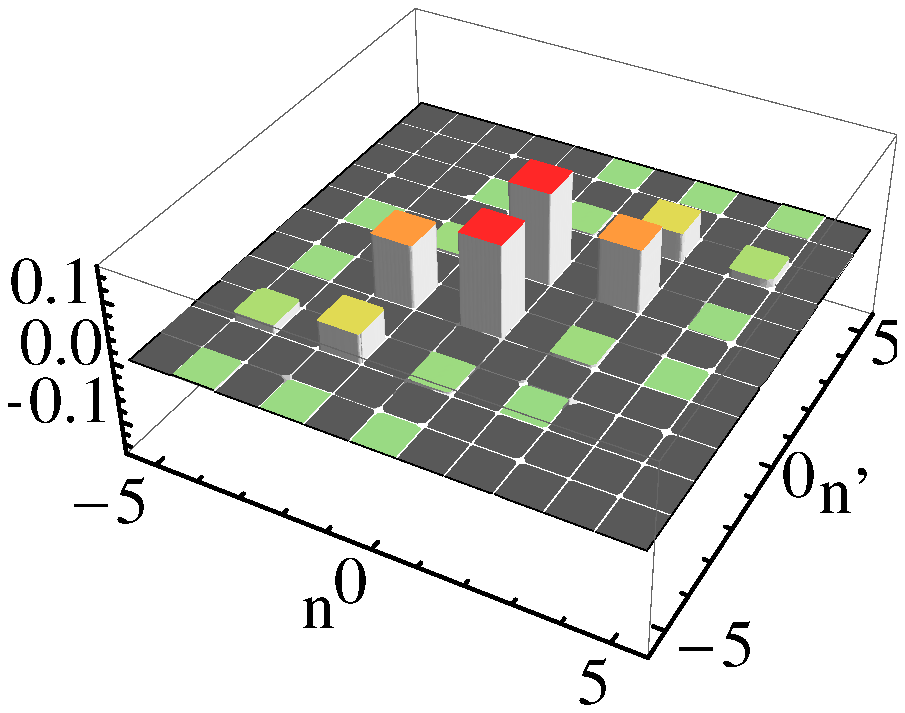}
\end{minipage}
\hfill
\begin{minipage}{.32\columnwidth}
\includegraphics*[width=\columnwidth]{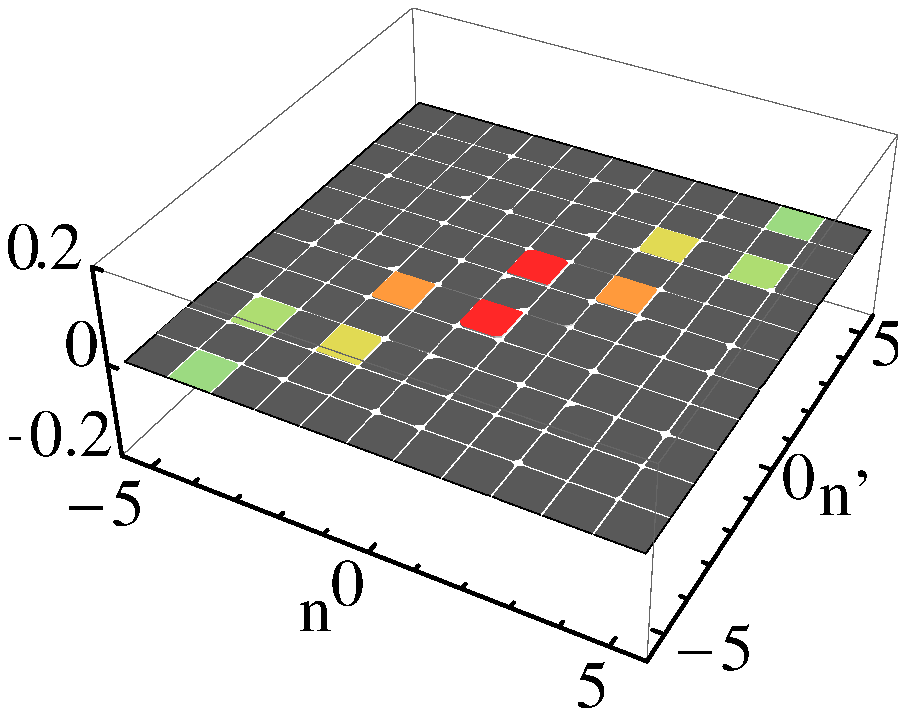}
\end{minipage}
\caption{(Color online)
Top: relaxation of the diagonal part of the  density
matrix  in the Fock basis (\ref{eq:density_matrix_10}) for $q=2$ and $N=10$  to the diagonal matrix
(\ref{eq:result_3}) as the noise is increased
from
$a_2 = 0$ (left) to $a_2 = 0.9$ (middle) and $a_2 = 2.9$ (right panel).
Bottom: off-diagonal part of  the density
matrix for $q=2$ and the same values of $a_2$, its vanishing indicates decoherence among the components of the macroscopic superposition.}
\label{fig:density_matrix_fock}
\end{figure}

Thus from (\ref{eq:density_matrix_10})-(\ref{casi2}) we obtain that
$\hat{\rho}_{\rm d} (t_q)  \rightarrow \hat{\rho} (\infty) $
when $a_qq \gg 1$ (phase relaxation) and $\hat{\rho}_{\text{od}} (t_q) \rightarrow 0$
when $a_q \gg 1$ (decoherence).
Hence, in the strong noise limit
the diagonal part of $\hat{\rho} (t)$ relaxes
to the steady state (\ref{eq:result_3}) and
the off-diagonal part
is washed away (see Fig.\ref{fig:density_matrix_fock}, right panels).
Remarkably, the decoherence factor in Eq.(\ref{eq:density_matrix_10}) does not depend on the atom number $N$. Also note the {\em different noise scales} relevant for
decoherence, $a_q$, and phase relaxation, $a_q q$. When increasing the noise intensity, 
$\hat{\rho}_{\rm d} (t_q)$ approaches $\hat{\rho}(\infty)$
{\em before} $\hat{\rho}_{\text{od}} (t_q)$ vanishes.
The higher the number of components  $q$ in the superposition, the more pronounced 
is this effect.
In fact, superpositions with higher $q$ are better protected against decoherence since they are formed at shorter times and $a(t)$ increases with time. 
Moreover, in the Markov regime $t_q \gg T_c$,
for a fixed noise intensity 
phase relaxation has a stronger effect on states with higher $q$, 
as illustrated in Fig.~\ref{fig:sketch_rot}. Indeed, by~(\ref{eq-a^2(t)}),
(\ref{eq:density_matrix_10}), and (\ref{casi}), the $n\not=n'$ matrix elements of 
$\hat{\rho}_{\rm d} (t_q)$ are   damped by a factor equal to or smaller than 
$e^{-a^2_q q^2/2} \simeq  \exp [ -(\pi q / \chi ) \int_0^\infty d \tau h(\tau)]$.
In the small-time regime $t_q \leq t_c$, 
all the $q$-component superpositions 
 relax to $\rho(\infty)$ at  the same ($q$-independant) 
noise   intensity (since  $a_q q$ is independent of $q$).
As a consequence of the distinct noise scales
for decoherence and phase relaxation, the BJJ does not turn
into a statistical mixture of phase states but relaxes directly to the mixture of 
Fock states (\ref{eq:result_3}).
This is illustrated in Fig.\ref{fig:density_matrix_fock} for the two-component superposition. 

Phase relaxation can be represented by the Husimi distribution of $\hat{\rho}_{\rm d}(t_q)$ \cite{Ferrini_2},
which for $q=2$, $N\gg 1$, and $a_2 \gg N^{-1/4}$ is given by
$Q_{\rm d}(\theta,\phi) =
\sum_{k=0}^1 \int d \phi ' f(\phi ', t_2 )
|\langle \theta,\phi |\frac{\pi}{2},\phi' +\pi k \rangle |^2/2
\simeq  Q_\infty(\theta) \Theta_3 ( - \phi-\pi \overline{\lambda}/(2\chi), e^{-2 a^2_2} )$
with  $\Theta_3$ the Theta function~\cite{Abramowitz}
and  $ Q_\infty (\theta)\simeq (\frac{1+\sin\theta}{2})^{N+1/2}/\sqrt{\pi N\sin\theta}$ the distribution
of the state
(\ref{eq:result_3}); $Q(\pi/2,\phi)$
is plotted for various values of $a_2$ in
Fig.\ref{fig:sketch_rot}. In the absence of noise
it shows  peaks at $\phi = 0$ and $\pi$, which correspond to the two coherent
states of the superposition. The peaks are smeared at increasing $a_2$, and finally at
$a_2 \gg 1 $ the Husimi distribution reaches the flat profile
$Q(\pi/2,\phi)=Q_\infty(\pi/2)$. 

An important consequence of the anomalous decoherence found in this model is that at
 intermediate noise strength associated to an already important phase relaxation, 
i.e., such that 
$\hat{\rho}_{\rm d} (t_q)$ is close to $\hat{\rho} (\infty)$ (e.g. for  $a_2=0.9$ in  Fig.\ref{fig:density_matrix_fock}),
 the system still  displays quantum correlations, which could be exploited in interferometry. 
These correlations can be quantified by computing the quantum Fisher information 
$F_Q[\hat{\rho}]$ associated to the $N$-particle density matrix $\hat{\rho}$~\cite{Caves94,Smerzi09}. The quantum Fisher information allows to estimate the best possible phase sensitivity $\Delta \theta$ in an interferometric scheme according to the generalized uncertainty principle $\Delta \theta \geq 1/( \sqrt{m} \sqrt{F_Q[\hat{\rho}]})$, $m$ being the number of measurements performed~\cite{Caves94}.
For separable states we have $F_Q[\hat{\rho}]\leq N$ and 
 $\Delta \theta \geq \Delta \theta_{\rm SN} = 1/\sqrt{m N}$.
Sub-shot noise sensitivities $\Delta\theta < \Delta \theta_{\rm SN}$
 can be achieved for states $\hat{\rho}$ satisfying $F_Q[\hat{\rho}]> N$, which is also 
a sufficient condition for multi-particle entanglement~\cite{Smerzi09}.
For $a_2 =0$, the two-component superposition $\hat{\rho}^{(0)}(t_2)$ has
$F_Q = N^2$~\cite{nota_fisher_opt}, corresponding to the Heisenberg limit 
$\Delta \theta=1/ \sqrt{m }N$.
For noise strength $a_2=0.9$ and $N = 10$ we find~\cite{nota_fisher_opt,future_work}
$F_Q [ \hat{\rho}(t_2) ] \simeq 58$, which is indeed still larger than 
$F_Q [ \hat{\rho}^{(0)}_{\rm d}(t_2)] = N$ and leads to 
a sensitivity gain $\Delta \theta/\Delta  \theta_{SN}$ of $-3.8$ db with respect to the use of separable states.
 In the limit $a_2 \gg 1$, 
$\hat{\rho}(t_2) \simeq \hat{\rho}(\infty)$ has a Fisher information $F_Q = N (N - 1)/(2N + 2) < N$~\cite{future_work}.


\section{VI. Summary and concluding remarks}
We have treated analytically the time evolution of a phase state
driven simultaneously by the atomic interactions and by a noise coupled to the
number imbalance operator $\hat{J}_z$. We have derived an exact expression for the density
matrix, as well as for the noise-induced decay
of the Ramsey visibility.
The effect of the noise on the creation
of macroscopic superpositions of phase states is to cause decoherence,
i.e., the vanishing of quantum correlations.
We have found that if the fluctuations of the atomic interactions
are negligible, decoherence is ``less efficient'' than phase relaxation, especially 
for superpositions with a large number of components. 
As a consequence, the states generated by the noisy dynamics
could  in principle lead to sub-shot noise  precision in interferometry.

The surprising fact that decoherence is not enhanced by increasing the 
atom number $N$ is specific to the noise 
considered. Indeed, such a noise is applied perpendicularly to the equator of 
the Bloch sphere where the phase states of the superpositions lay. As a result, the noise
 is insensitive to the separation between these states, which scales with $N$.
However, such superpositions are very fragile under a noise applied parallel to the
 equatorial plane, which resolves the separation between the components. 
This yields an indication as to which classical noise to reduce to preserve the 
coherence in superpositions of the phase states: this is the noise in directions
 parallel to this plane.
For example, stochastic fluctuations on the tunnel amplitude $K$ give rise  to rapid 
decoherence of the macroscopic superposition 
$(| \alpha=1\rangle + e^{i \gamma} | \alpha =-1\rangle)/\sqrt{2}$ at a rate increasing with the atom number, without inducing relaxation.

\section{ACKNOWLEDGMENTS}

We thank C. Gross and M. Oberthaler for useful discussions on experimental issues and a careful reading of the manuscript.
We acknowledge financial support from CNRS, the MIDAS project, and
the project ANR-09-BLAN-0098-01.



\begin{thebibliography}{33}
\bibitem{nota_fisher_opt} We compute the maximal Fisher information over all
directions of angular momentum $\bold J$ as $F_Q = 4 \lambda_{\rm max}$,
 where $\lambda_{\rm max}$ is 
the largest eigenvalue of the $3\times 3$ 
matrix $\Gamma_{i,j} = \sum_{l,m} \frac{(p_l-p_m)^2}{p_l+p_m} 
\langle l| \hat{J}_i|m \rangle 
\langle m| \hat{J}_j|l \rangle/2$ and $|l\rangle$ the normalized eigenvectors 
of $\hat{\rho}$ with eigenvalues $p_l$, see \cite{Smerzi09}.
\end{thebibliography}
\end{document}